\documentclass[a4paper]{aip-cp}

\usepackage[numbers]{natbib}
\usepackage{rotating}

\usepackage[T1]{fontenc}
\usepackage{textcomp}   % provides proper micro- and degree-symbol
\usepackage{gensymb}
\usepackage[latin1]{inputenc}
\usepackage[english]{babel}
\usepackage[dvipsnames,pdftex]{xcolor}
\usepackage{amsbsy} 
\usepackage[caption=false]{subfig}
\usepackage{multicol}
\usepackage{multirow}
\usepackage[tight,nice]{units}
\usepackage{siunitx}%added by Martin
\newlength{\bildtitel}
\newcommand\REVIEW[1]{\message{LaTeX Warning: \noexpand untreated nEDM-REVIEW command in \jobname .tex: l\the\inputlineno}}% for publication
\newcommand{\mbf}{\boldsymbol}
\renewcommand{\parallel}{\uparrow\!\uparrow}
\renewcommand{\nparallel}{\uparrow\!\downarrow}

\newcommand{\diff}[1]{\operatorname{d}\ifthenelse{\equal{#1}{}}{\,}{\!#1}}

\newcommand{\pow}[2]{\ensuremath{#1\!\times\!10^{#2}}}

\newcommand{\ecm}{\ensuremath{\si{\elementarycharge}\!\cdot\!\cm}}
\newcommand{\dn}{\ensuremath{d_\text{n}}}

\def\Bra#1{\left\langle#1\right|}
\def\Ket#1{\left|#1\right\rangle}

\newcommand{\stateup}{\ensuremath{\Ket{\uparrow}}}

\newcommand{\cm}{\ensuremath{\mathrm{cm}}}

\newcommand{\tHe}{\ensuremath{{}^3\mathrm{He}}}

\newcommand{\magHg}{\ensuremath{{}^{199}\text{Hg}}}

\newcommand{\magXe}{\ensuremath{{}^{129}\mathrm{Xe}}}

\addto\captionsenglish{\renewcommand{\figurename}{\bf FIGURE}}
\addto\captionsenglish{\renewcommand{\tablename}{\bf TABLE}}

\begin{document}

% Title portion
\title{The quest for an electric dipole moment of the neutron}

\author[aff1]{P.~Schmidt-Wellenburg\corref{cor1}}

\affil[aff1]{Paul Scherrer Institute, 5232 Villigen, Switzerland}
\corresp[cor1]{Corresponding author: philipp.schmidt-wellenburg@psi.ch}

\maketitle

\begin{abstract}
Until this day no electric dipole moment of the neutron (nEDM) has been observed. Why it is so vanishing small, escaping detection in the last 50 years, is not easy to explain. In general it is considered as the most sensitive probe for the violation of the combined symmetry of charge and parity (CP). A discovery could shed light on the poorly understood matter/anti-matter asymmetry of the universe. As nucleon it might one day help to distinguish different sources of CP-violation in combination with measurements of the electron and diamagnetic EDMs.
This proceedings articles presents an overview of the most important concepts in searches for an nEDM and presents a brief overview of the world wide efforts.
\end{abstract}

% Head 1
\section{INTRODUCTION}
With the discovery of the Higgs-boson\,\cite{Aad2012PLB,Chatrchyan2012PLB} particle physics at high energy colliders celebrated its latest success in an exceptional series of discoveries which all were confirming the standard model of particle physics (SM). The non-observation or lack of any signal indicating new physics in collider experiments in combination with several astrophysical observations (i.e.\ dark matter, neutrino-oscillation) nurtured vivid interest in high precision physics at low energies\,\cite{Raidal2008} in the last years. One such high precision search for new physics is the quest for a electric dipole moment of the neutron. While in the recent past remarkable progress were made for the electron EDM using the enormous electric field inside a ThO molecule\,\cite{Baron2014Science} and for the diamagnetic \magHg-EDM\,\cite{Graner2016arXiv}, it remained relatively silent around the neutron. The latest result for the neutron EDM $|\dn|\!<\!\unit[\pow{2.9}{-26}]{\ecm}$ is dating back to 2006\,\cite{Baker2006} was recently re-analyzed\,\cite{Pendlebury2015PRD} (now $|\dn|\!<\!\unit[\pow{3}{-26}]{\ecm}$) confirming the original result. Different EDM limits are summarized in Tab.\,\ref{tab:EDMs}, while the interested reader is referred to Ref.\,\cite{Chupp2015PR} for a detailed overview of the interplay of different EDMs and theory. 

\begin{table}%
\centering
\begin{tabular}{rcl}
\hline
EDM & limit & C.L. \\ 
\hline
ThO & $d_e<\pow{8.7}{-29}\ecm$ &90\%\\
\magHg & $d_{\rm Hg}<\pow{7.4}{-30}\ecm$ &95\% \\
neutron & $\dn <\pow{3.0}{-26}\ecm$ &90\%\\
\hline
\end{tabular}
\caption{Most relevant experimental limits on electric dipole moments.}
\label{tab:EDMs}
\end{table}

First searches, starting in the 1950s\,\cite{Purcell1950PR,Smith1957PR}, for an nEDM were undertaken using thermal, later cold neutron beams from reactors. The last beam experiment\,\cite{Dress1977} published a limit of $\dn\!<\!\unit[\pow{3}{-24}]{\ecm}$ (C.L.90\%), and was limited by the velocity dependent $\boldsymbol{v\!\times\!E}$ systematic effect. At the beginning of that decade first experiments were proposed using ultracold neutrons\,\cite{Lushchikov1969JETPL,Shapiro1970SPU} with the first results published at the end of the 70s\,\cite{Altarev1980NuPhA}.
Today several competing collaboration around the world pursue new experiments to improve the limit on the nEDM by up to 2 orders of magnitude in the next decade. This proceeding article reviews the most relevant aspects concerning the quest for an nEDM describing generically the typical techniques in use. For a complete review I would like to refer the reader to the article by Golub and Lamoreaux\,\cite{Lamoreaux2009}.

\section{THE nEDM IN THE STANDARD MODEL AND BEYOND}
In fundamental physics, symmetries and their violation have been an important concept and guidelines for theories and models. Already in 1918 Emmy Noether\,\cite{Noether1918} showed that the conversation of energy, momentum and angular momentum can be directly derived from symmetries under time reversal, translation or rotation, respectively. 
In modern particle physics three discrete symmetries: charge conjugation ({\it C}), parity inversion ({\it P}), and time reversal ({\it T}) play an outstanding role in our understanding of nature. 
All of these are individually violated in the weak sector of the SM, further the combined symmetry of charge and parity ({\it CP}) is violated in decays of {\it K} and \textit{B} mesons. 
Although all necessary ingredients for baryon-genesis (Sakharov criteria\,\cite{Sakharov1991}) in the early universe exist in the SM, the prediction of the baryon asymmetry of the Universe (BAU) $\eta\approx 1\!\times\!10^{-18}$ from the SM weak sector\,\cite{Riotto1999ARNPS} falls eight orders of magnitude short when compared to values derived from the measurement of the microwave background of the universe\,\cite{Dine2012} $\eta=6.1^{+0.3}_{-0.2}\!\times\!10^{-10}$, or from the abundance of light elements produced in primordial nucleosynthesis\,\cite{Cyburt2008JCAP} with $5.1\!\times\!10^{-10}<\eta<6.7\!\times\!10^{-10}$. One essential ingredient for baryon-genesis is the violation of {\it CP}-symmetry\,\cite{Sakharov1991}. 

\begin{figure}%
	\centering
	\includegraphics[width=0.3\columnwidth]{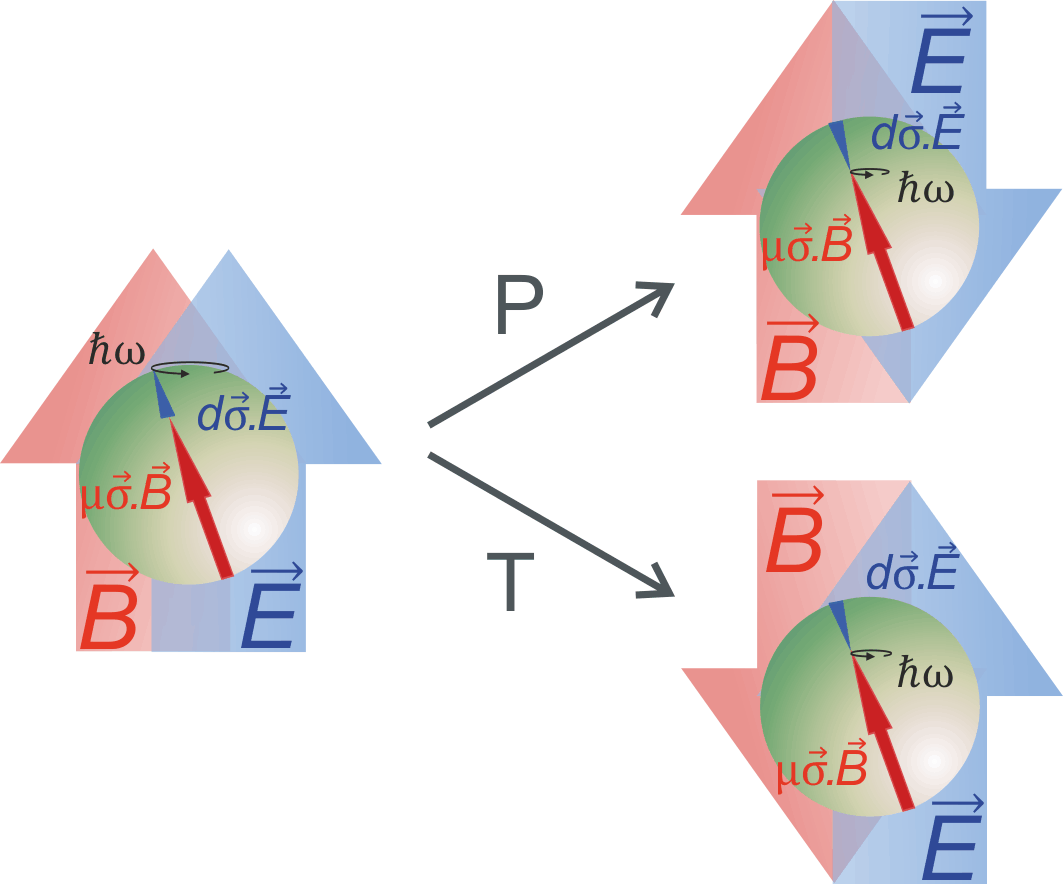}%
	\caption{Cartoon of {\it P} and {\it T}-violation of a nEDM in an electric and magnetic field.}%
	\label{fig:CPViolCartoon}%
\end{figure}

The existence of an electric dipole moment of a neutron would manifest a new source of {\it CP}-violation (CPV)\@. If a neutron would have an ``electric dipole moment \dn{} then, as
any vector operator in quantum mechanics, it is connected to
the spin operator as $ \dn= \delta_\text{n} \mathbf{j}/j\hbar$, or, for $j= 1/2$ as $\dn= \delta_\text{n}\boldsymbol{\sigma}$,
where \dn{} gives the size of the EDM, usually in units of \ecm''\cite{Dubbers2011}. In the non-relativistic limit, the interaction Hamiltonian can be written as:

\begin{equation}
	H = -\frac{\hbar}{2}(\delta_\text{n}\boldsymbol{\sigma\!\cdot\!E} + \gamma_\text{n} \boldsymbol{\sigma\!\cdot\!B}),
	\label{eq:hamiltonian}
\end{equation}

\noindent where $\delta_\text{n}$ and $\gamma_\text{n}$ can be interpreted as scalar coupling strengths of the neutron spin to the electric and magnetic field. The relative sign of the two dipole coupling strengths is not yet defined as no electric dipole moment has yet been discovered. The magnetic coupling strength is nothing else than the gyromagnetic ratio of the neutron $\gamma_\text{n}/(2\pi)=\SI{-29.1646943(69)}{MHz\per\tesla}$\,\cite{Greene1978}, which is the ratio of the magnetic moment of the neutron $\boldsymbol{\mu_\text{n}}$ to its angular momentum $\sigma = \hbar/2$. Similar one can introduce a gyroelectric ratio in combination with the electric dipole moment. Equation\,(\ref{eq:hamiltonian}) and Fig.\,\ref{fig:CPViolCartoon} demonstrate that the eigenstates of the Hamiltonian change when applying either a {\it C} or {\it P}- transformation to the Hamiltonian, indicating the violation of {\it C} and {\it P}-symmetry. The CPT-theorem (see standard text books on field theory, e.g.\,\cite{Maggiore2005Book}) is fundamental to any modern quantum field theory and states, that any locally Lorentz-covariant field theory of a point like particle is CPT invariant. This indicates that the observation of \dn{} would not only indicate time reversal symmetry breaking but also CPV and might help to explain the observed BAU\@. 

\subsection{The Standard Model Prediction}
In the SM two sources of CPV exist, for a detailed discussion see also the review by Pospelov and Ritz\,\cite{Pospelov2005}: {\it i})~In the weak interaction the weak mass eigenstates of the quarks are not identical to the flavor eigenstates. Both eigenstates are interconnected via the Kobayashi-Maskawa matrix $V_\text{KM}$ which has one single phase $\delta$ which induces the observed CPV in the K and B meson decays.
	{\it ii})~The second source is the QCD vacuum polarization term, the only CP-odd term of dimension four in the SM QCD Lagrangian.

\begin{figure}%
\centering
\includegraphics[width = 0.25\columnwidth]{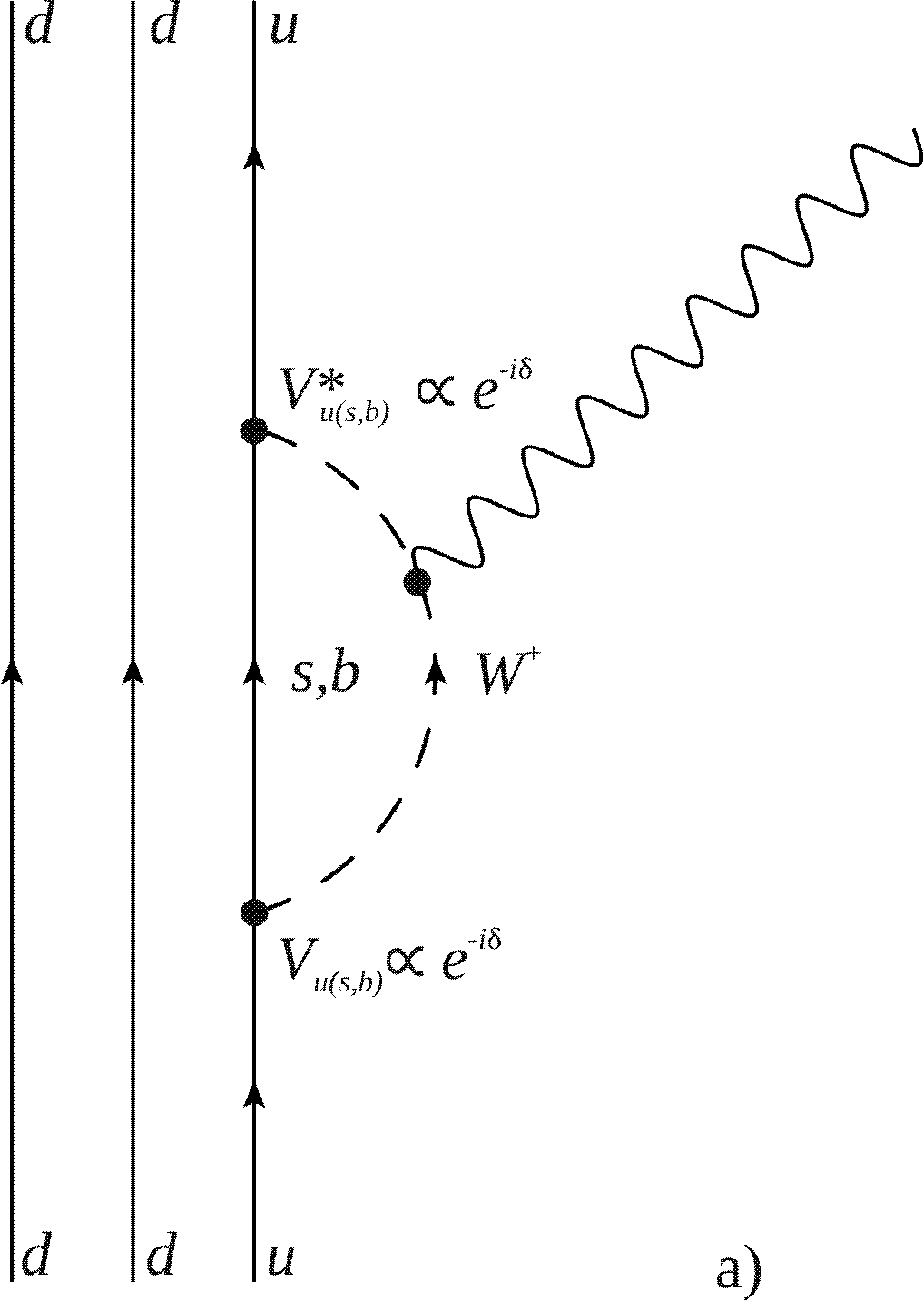}%
\hspace{3cm}
\includegraphics[width = 0.4\columnwidth]{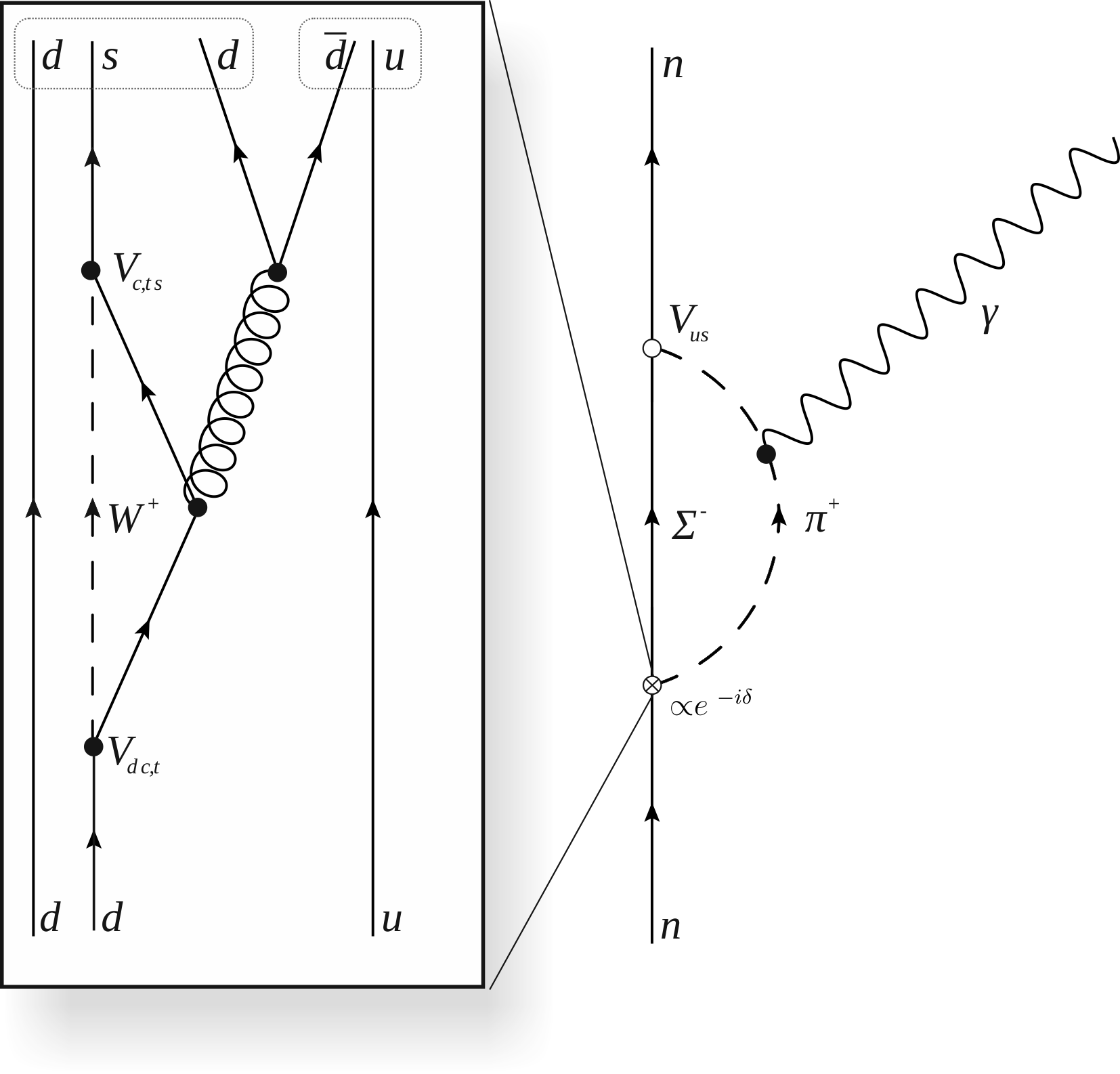}
\caption{Diagrams which involve the CPV phase of the $V_{\rm KM}$-matrix. a)~First loop contribution does not give  rise to an EDM, as the phase from the first vertex cancels in the complex conjugate second vertex. b)~This is the largest SM contribution generated by the CPV phase of the $V_{\rm KM}$-matrix. The crossed vertex, shown as insert, is a four-quark vertex known as strong penguin diagram in which not all phase contributions cancel.}%
\label{fig:WeakDiagramloops}%
\end{figure}

The Kobayashi-Maskawa matrix $V_\text{KM}$ can be written as
\begin{equation}
	V_\text{KM} = \left(\begin{array}{ccc} 
											c_{12}c_{13} & s_{12}c_{13} & s_{12}e^{-i\delta} \\
											-s_{12}c_{23}-c_{12}s_{23}s_{13}e^{-i\delta} & c_{12}c_{23}-s_{12}s_{23}s_{13}e^{-i\delta} & s_{23}\\
								s_{12}s_{23} - c_{13}c_{23}c_{13}e^{-i\delta}			&  - c_{12}s_{23}s_{13}e^{-i\delta}& c_{23}c_{13}\\
								\end{array}\right)
\label{eq:VKM}
\end{equation}
\noindent where $c_{ij} = \cos\theta_{ij}$, $s_{ij} = \sin\theta_{ij}$ and $\delta \approx \unit[1.20]{rad}$ is the CPV phase. It is impossible to write down a tree level diagram generating an electric dipole interaction of one quark of the neutron with the electric field. At the one-loop level, shown in Fig.\,\ref{fig:WeakDiagramloops}a), any phase term of a $V_{ij}$ element at one vertex will be canceled by the complex conjugated phase term at the second vertex $V_{ij}^{\ast}$. Shabalin\,\cite{Shabalin1983} showed that even all two-loop level contributions to an nEDM cancel. The largest SM contribution is at the three-loop level via a strong penguin diagram\,\cite{Khriplovich1982PL} (see Fig.\,\ref{fig:WeakDiagramloops}b) which amounts to an approximate $\dn{}^\text{KM}$ of \unit[\pow{1}{-32}]{\ecm}\,\cite{Pospelov2005}, well below current and most probable all future experimental sensitivities.

An additional source of CPV in the SM is the vacuum term of the QCD-Lagrangian:

\begin{equation}
	\mathcal{L}_\text{QCD}^\text{CPV}=\frac{g_s^2}{32\pi^2}\overline{\theta}G^a_{\mu\nu}\widetilde{G}^{\mu\nu,a},
\label{eq:QCDDim4}
\end{equation}

\noindent the only CPV dim-4 operator, where $g_s$ is the coupling constant of the strong interaction, $ \overline{\theta}$ is a phase which also includes the CPV phase of the weak interaction and $G^a_{\mu\nu}$ is the gluon field tensor. The structure of the gluon field tensor times its dual corresponds in electro-magnetism to a scalar product of $\mbf{E}\cdot\mbf{B}$ which is odd under {\it P} and {\it T} reversal. From a dimensional analysis\,\cite{Khriplovich1997Book} one can estimate the size of an nEDM generated by this term:

\begin{equation}
	\dn^\text{QCD} \approx \overline{\theta}\cdot\unit[\pow{1}{-16}]{\ecm}.
\label{eq:QCDThetaApprox}
\end{equation}

\noindent Hence, the current experimental limit on an nEDM is also a limit on $\overline{\theta} < 10^{-10}$. This is astonishing, as $\overline{\theta}$ is a phase which in principle could acquire any value between $0$ and $2\pi$. It is considered as unnatural that the value is so tiny. Possible solutions to this ``strong {\it CP} problem'' is either at least one mass-less quark (unlikely), or a mechanism proposed by Peccei and Quinn\,\cite{Peccei1977,Wilczek1978,Weinberg1978}, which gives rise to the axion, a Nambu-Goldstone boson. For searches for the axion and axion like particles see, e.g., Ref.\,\cite{Baer2015}. However, if an nEDM would be found further measurements of EDMs (proton, electron,...) will be necessary to distinguish the CPV source(s) and to explain the role of the tiny $\overline{\theta}$-term.

\subsection{Generic Sensitivity of an nEDM to Physics Beyond the SM}
In the scope of this proceeding it is impossible to cover the variety of different BSM scenario which provide viable sources of CPV, the interested reader may refer to the reviews Refs\,\cite{Jaeckel2012JHEP,Engel2013PPNP} for a more comprehensive summary. Neglecting for a moment the contribution to an nEDM from the un-natural small $\overline{\theta}$ parameter, it is clear that any observed $\dn>\unit[10^{-30}]{\ecm}$ is most likely explained by new physics. Generically most BSMs provide several CPV phases and new particles which could already contribute at the one-loop quark level to an observable nEDM\@.  A typical order of magnitude analysis from super symmetric (SUSY) models (e.g.\,\cite{Abel2006JHEP}) gives:

\begin{equation}
		\dn \sim \left(\frac{\unit[300]{GeV}}{\Lambda_{\rm SUSY}}\right)^2\sin\phi_{\rm CP}\times10^{-24}\ecm,
\label{eq:BSMapprox}
\end{equation}

\noindent where $\phi_{\rm CP}$ represents the relevant possible CPV phases of the model and $\Lambda_{\rm SUSY}$ is the SUSY mass scale. The current experimental limit already implies that models either have to be considerable fine tuned to have a small $\phi_{\rm CP}$ or to suppress 1-loop contributions, or that the SUSY-scale is considerable above the weak-scale in the range of some TeV\@. This SUSY approach is generalized in Ref.\,\cite{Engel2013PPNP} and results in a similar model independent constraint for a general BSM scale.

\section{Experimental Techniques}
Experiments searching for the electric dipole moment of the neutron essentially try to measure the precession frequency of the neutron in a strong electric field $\omega_\text{E} = \delta_\text{n}\mbf{E}$. The current upper limit of $\dn \leq \unit[\pow{3}{-26}]{\ecm}$ (90\% C.L.) indicates that it will be necessary to measure a frequency below $2\pi\omega_\text{E}=\SI{24}{nHz}$ for an electric field of $E=\SI{10}{kV}$. Any magnetic field larger than \SI{0.83}{fT}, a field too small for even the best magnetically shielded rooms on Earth, would lead to a similar or even larger Larmor precession frequency of the neutrons. Thus, it seems impossible to directly measure the effect of an nEDM exposed ``only'' to an electric field. Instead the neutron is exposed in addition to a well controlled magnetic field $\boldsymbol{B}$. By taking the difference of two Larmor frequencies measured in configuration where the electric field is parallel ($\omega^{\parallel}$) or anti-parallel ($\omega^{\nparallel}$) to the magnetic field:

\begin{eqnarray}
	\hbar\omega^{\parallel} &= & 2\left|\boldsymbol{\mu_\text{n}\!\cdot\!B^{\parallel}}+\boldsymbol{\dn\!\cdot\!E^{\parallel}}\right| \nonumber\\
	\hbar\omega^{\nparallel} &= & 2\left|\boldsymbol{\mu_\text{n}\!\cdot\!B^{\nparallel} }-\boldsymbol{\dn\!\cdot\!E^{\nparallel} }\right| \nonumber\\
	\dn &= & \frac{\hbar\left(\omega^{\parallel}-\omega^{\nparallel} \right)-2\mu_\text{n}\left(B^{\parallel}-B^{\nparallel}\right)}{2\left(E^{\parallel}-E^{\nparallel}\right)}
\label{eq:DiffConfig}
\end{eqnarray}

\noindent In general these two measurement are either made in two adjacent volumes with opposite electric fields ($E^{\parallel}=-E^{\nparallel}$) inside the same magnetic field ($B^{\parallel}-B^{\nparallel} = 0$), or by measuring first one configuration, then changing the polarity of the electric field from $E^{\parallel}$ to $E^{\nparallel}=- E^{\parallel}$ and measure again. In the first case it will be of paramount importance to make sure that the two spatial separated measurement have the same magnetic field configuration (no or small magnetic-field gradients), while in the second case it is essential to make sure that the magnetic field is stable in time. Both approaches are currently used or in discussion for searches of an nEDM\@.

\subsection{Ramsey's technique of separated oscillating fields}
\begin{figure}%
\centering

\includegraphics[width=0.45\columnwidth]{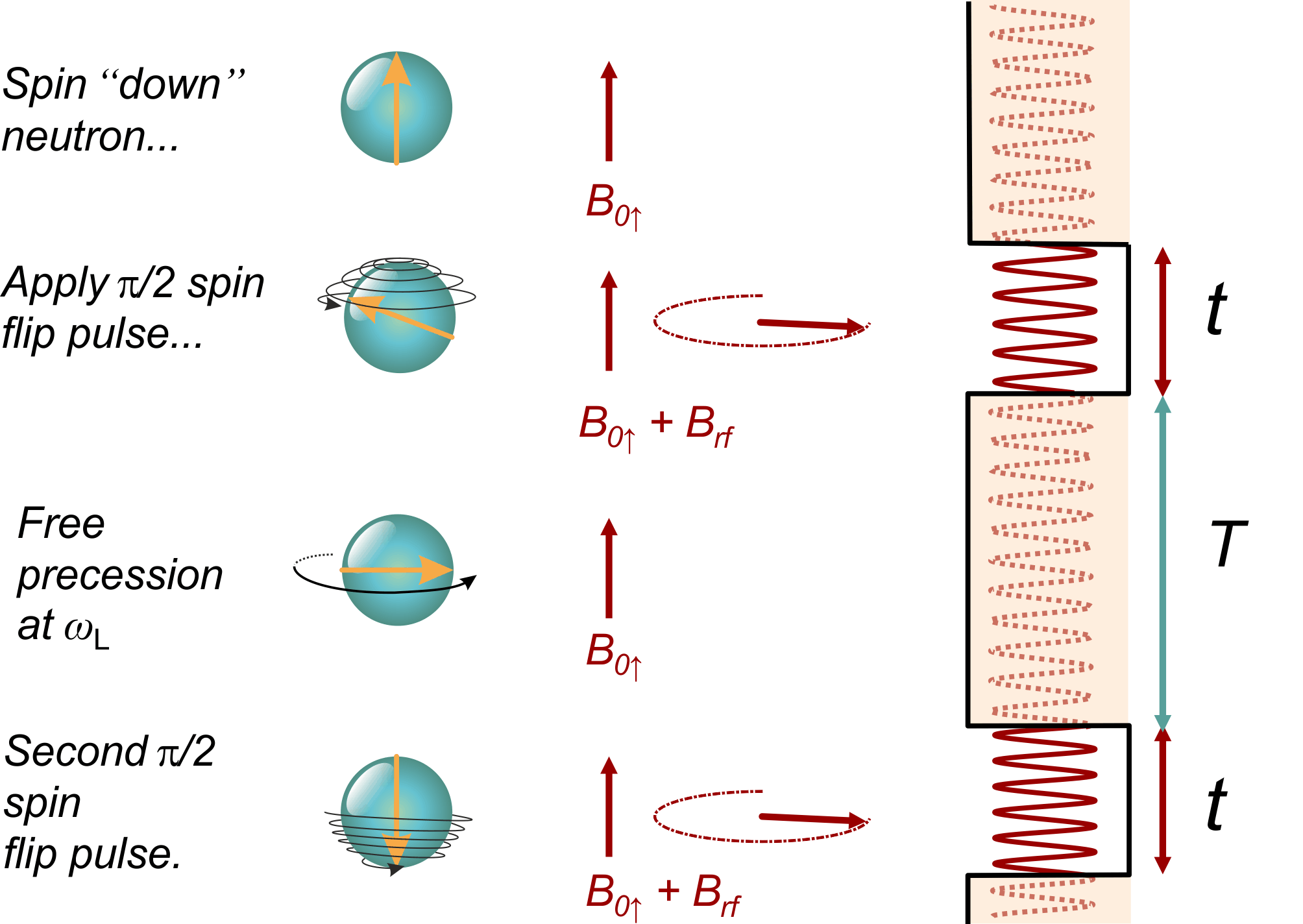}
\hspace{0.04\columnwidth}

\includegraphics[width=0.43\columnwidth]{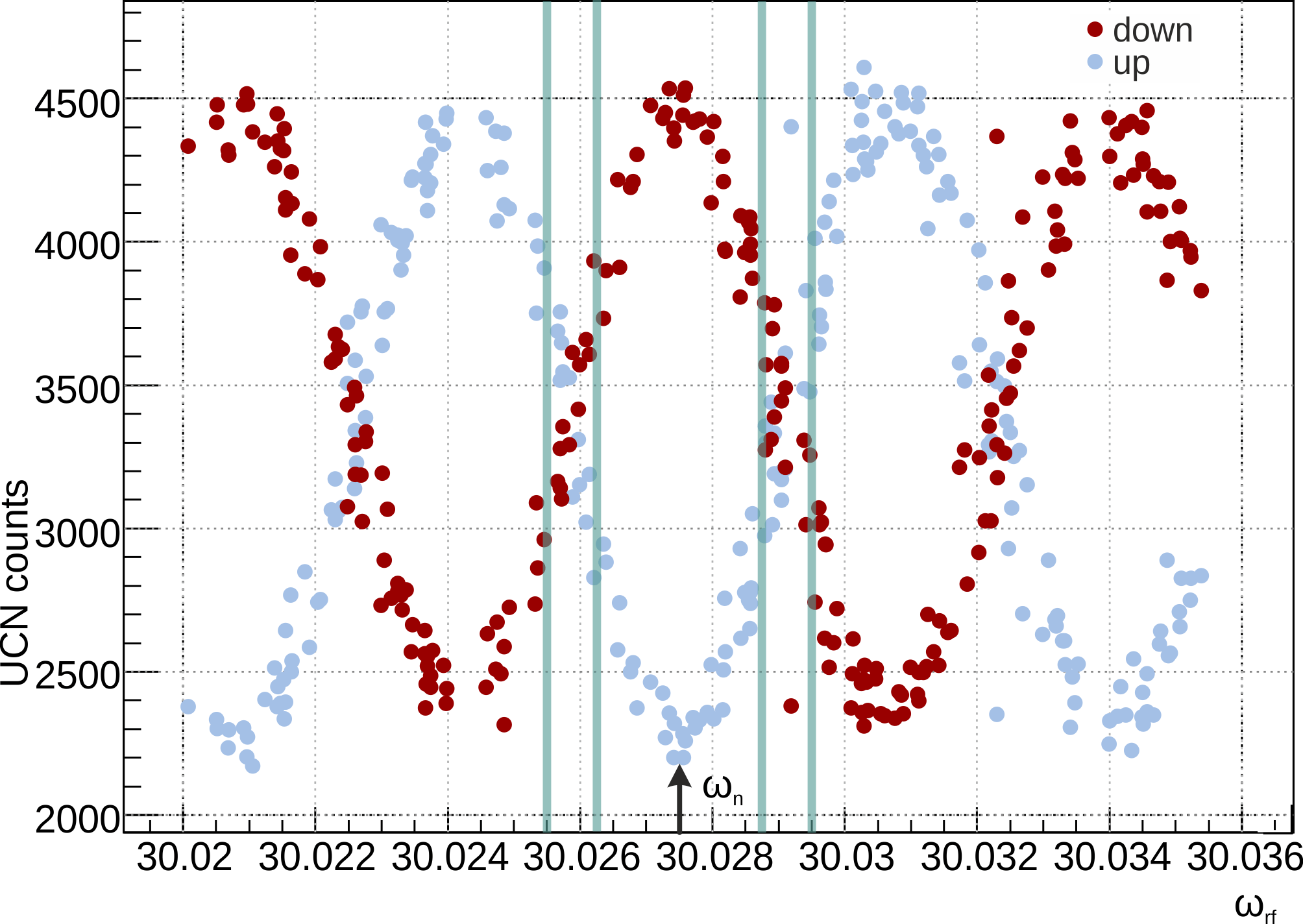}
\caption{Ramsey's technique of separated oscillating fields. The scheme of the method (left) and the data plot (right) are explained in detail in the text. Blue points are UCN counted with spin up $N_\text{u}$, while red points are with spin down $N_\text{d}$ (data from the PSI-nEDM collaboration\,\cite{Baker2011}. The width at half height $\Delta \nu$ of the central fringe is approximately $1/\left(2T\right)$, the four vertical lines indicate the working points. }%
\label{fig:RamseyMethod}%
\end{figure}

More than half a century ago Ramsey improved Rabi's resonant frequency technique to measure energy eigenstates of quantum mechanical systems by introducing a free precession period between two spin-flipping pulses\,\cite{Ramsey1950PR}.
Figure\,\ref{fig:RamseyMethod}a) illustrates this technique while a typical resonance scan is shown in Fig.\,\ref{fig:RamseyMethod}b). The initial state is a fully polarized, i.e.\ \stateup, ensemble of neutrons exposed to a magnetic field $B_0$. A first rotational oscillating magnetic-field pulse $B_1\cos\left(\omega_\text{rf}t\right)$, perpendicular to $B_0$  tips the spins into the plane orthogonal to the main magnetic field. The neutron spin then precesses freely with their Larmor frequency $\omega_0$ for a duration $T$, accumulating a phase $\phi = \gamma_n B T$, before a second pulse $B_1\cos\left(\omega_\text{rf}t\right)$ in phase with the first is again applied to the neutron ensemble. The essential idea, is to compare the phase $\phi$ with $\omega_{\rm rf}T$. If they are identical then $B=\omega_{\rm rf}/\gamma_n$. 

The probability to detect a neutron with a final spin state identical to its initial spine state, i.e.\ \stateup, is (see equation~(A.11) in Ref.\,\cite{Piegsa2009NIMA}):

\begin{equation}
	\mathcal{P}(T,\omega_\text{rf})=\left|\Bra{\uparrow}U(T,\omega_\text{rf})\Ket{\uparrow}\right|^2 =1-\frac{4\omega_1^2}{\Omega^2}\sin^2\frac{\Omega t_{\pi/2}}{2}
	\left[\frac{\Delta}{\Omega}\sin\frac{\Omega  t_{\pi/2}}{2}\sin\frac{T\Delta}{2}-\right. 
	\left.\cos\frac{\Omega  t_{\pi/2}}{2}\cos\frac{T\Delta}{2}\right]^2,
\label{eq:RamseyResonance}
\end{equation}
\noindent where $U(T,\omega_\text{rf})$ is the time evolutions operator describing the pulse sequence, $\omega_1\!=\!-\gamma_\text{n}B_1$, $\Delta \!=\!\omega_\text{rf}-\omega_0$, and  $\Omega\!=\!\sqrt{\Delta^2+\omega_1^2}$. In an optimized frequency the spin-flipping pulses have just exactly enough power to tip the spins by $\pi/2$, hence, the pulse length and field power fulfill the condition $\gamma_\text{n}B_1t_{\pi/2} =\pi/2$. In this case and in the central fringe range ($\Delta \ll \omega_1$) equation\,(\ref{eq:RamseyResonance}) simplifies to:

\begin{eqnarray}
		\mathcal{P}(T,\omega_\text{rf}) &=&1-4\sin^4\frac{\pi}{4}
	\left[\frac{\Delta}{\Omega}\sin\frac{T\Delta}{2}-\cos\frac{T\Delta}{2}\right]^2 \nonumber \\
	\mathcal{P}(T,\omega_\text{rf}) &\approx& 1- \cos^2\frac{T\Delta}{2} \nonumber \\
	\mathcal{P}(T,\omega_\text{rf}) &=& \frac{1}{2}\left(1 -\cos\left(T\Delta\right)\right).
\label{eq:CosineApproximation}
\end{eqnarray}

\noindent In a real measurement with $N$ neutrons inside a large magnetic field region this becomes:

\begin{equation}
	N^{\uparrow} = \frac{N}{2}\left\{1 - \alpha(T)\cos\left[ \left(\omega_\text{rf} - \gamma_\text{n}B_0\right)\cdot\left(T+\frac{4t_{\pi/2}}{\pi}\right)\right]\right\},
\label{eq:UCNRamseyFormula}
\end{equation}
\noindent where $\alpha(T)$ is the visibility of the central fringe taking into account all depolarization effects\,\cite{Afach2015PRD}. The term  ${4t_{\pi/2}}/{\pi}$ is necessary to account for field inhomogeneities of $B_1$ and $B_0$ which become relevant when the pulse length $t_{\pi/2}$ is finite\,\cite{Slichter1990book}. 

A Ramsey interference pattern as shown in Fig.\,\ref{fig:RamseyMethod}b), can be recorded by scanning $\omega_\text{rf}$ while keeping all other conditions constant. The magnetic field $B$ measured by the neutrons then fulfills the resonance condition of the central fringe with $\omega_\text{rf} = \gamma_\text{n}B$.
This procedure is slightly optimized in searches for an nEDM, by only measuring at four points with highest sensitivity, the working points.
The neutron Larmor frequency is then obtained by fitting expression\,(\ref{eq:UCNRamseyFormula}) to the data.
Separate fits are performed for different electric field and magnetic field configurations.
Taking the difference of these Larmor frequencies then give access to the electric dipole moment:

\begin{equation}
		\dn = \frac{\hbar (\omega_0^{\parallel}-\omega_0^{\nparallel})}{2(E^{\parallel} - E^{\nparallel})} = \frac{\hbar \Delta\omega}{4E},
\label{eq:nEDM-simple}
\end{equation}

\noindent using equation\,(\ref{eq:DiffConfig}) and assuming no differences in the magnetic field and $E=E^{\parallel} = -E^{\nparallel}$. The statistical  sensitivity with which a frequency can be measured can be deduced from equation\,(\ref{eq:UCNRamseyFormula}):

\begin{equation}
	\sigma\left(\omega_0\right) \approx \frac{1}{\alpha T \sqrt{\langle N\rangle}},
\label{eq:SensitivityFrequency}
\end{equation}

\noindent where $\langle N\rangle$ is the average total number detected neutrons. This translates into a statistical sensitivity for the detection of an nEDM of:

\begin{equation}
	\sigma\left(\dn\right) \approx \frac{\hbar\sqrt{2}}{4\alpha T E \sqrt{\langle N\rangle/2}}
	= \frac{\hbar}{2\alpha T E \sqrt{\langle N\rangle}},
\label{eq:SensitivityEDM}
\end{equation}

\noindent  using equation\,(\ref{eq:nEDM-simple}). The additional factor $1/\!\sqrt{2}$ in the denominator accounts for the fact that it needs two measurements to take a difference.

\subsection{Ultracold neutrons}
Starring at equation\,(\ref{eq:SensitivityEDM}) immediately makes it obvious that $T\!\sqrt{N}$ needs to be maximized. One possible solution, proposed already in 1960by Shapiro\,\cite{Lushchikov1969JETPL,Shapiro1970SPU}, is to increase $T$ by using neutrons which can be stored within vacuum chambers made of adequate materials. Such ultracold neutrons (UCN) are reflected under any incident angle by the neutron optical potential

\begin{equation}
	V = \frac{2\pi\hbar^2}{m_\text{n}}\mathcal{N}b,
\label{eq:NeutronOptPot}
\end{equation}

\noindent where $m_\text{n}$ is the neutron mass, $\mathcal{N}$ the nucleon density and $b$ the neutron scattering of a given material. Well suited materials for storage are, e.g., $^{58}$Ni, diamond like carbon (DLC), NiMo(85/15), or BeO/Be which all have neutron optical potentials in the range  \SIrange{200}{350}{neV}. These potentials also define the maximum kinetic energy of UCN\@. Which is in the same range as the gravitational potential energy of neutrons $mgh\approx h
\cdot\unit[1.025]{neV/cm}$. Further, strong magnetic fields can be used to polarize or store UCN, as the magnetic potential energy $\boldsymbol{\mu}\cdot\boldsymbol{B} \approx \pm \mbf{B}\cdot\unit[60]{neV/T}$. This means, a magnetic field of SI{5}{T} creates a potential barrier of \SI{300}{neV} for one spin state while the other spin state is attracted. For a complete review of UCN physics please refer to the books\,\cite{Golub1991,Ignatovich1990}.

In current nEDM-experiments connected to existing UCN-sources up to 15\,000 UCN are counted after a storage of \unit[180]{s}, which results in a figure of merit of $T\sqrt{N} =\unit[27000]{s}$, and a statistical sensitivity of $\sigma\left(\dn\right) \approx \unit[\pow{1}{-25}]{\ecm}$ per day of measurement. Further progress will be possible only by significantly increasing the number of ultracold neutrons.

Any cold neutron source also has a significant amount of UCN in the low energy tail of the Maxwell distribution. Extraction through aluminum windows and long guides significantly reduces the amount of available UCN outside the biological protection. These problems were beautifully circumvented by the conception and design of a phase-space converter, the UCN-turbine (instrument PF2) at ILL which Doppler-shifts very cold neutron to the UCN regime\,\cite{Steyerl1975NIM,Steyerl1986PLA} which is until today the working horse and benchmark in UCN-physics. All other current and next generation UCN sources\,\cite{Trinks2000NIMA,Saunders2013RSI,Lauss2014,Piegsa2014,Serebrov2014TPL,Masuda2015JPS} are based on the superthermal concept proposed by Pendlebury and Golub already in 1975\,\cite{Golub1975,Golub1977}.

The principal idea is to use collective excitations of the conversion medium to down-scatter neutrons from higher energies to the UCN energy regime. For this process two materials are of main interest: superfluid helium (He-II), using the phonon-roton excitations with a relative feeble production rate but profiting from a zero absorption cross-section\,\cite{Korobkina2002,Schmidt-Wellenburg2009}, and solid deuterium (sD$_2$) which has a broad range of excitations leading to a high conversion rate while the finite absorption cross-section reduces the effective layer from which a UCN can escape the material\,\cite{Atchison2007PRL}. Both method can be adapted in such a way that in principle it should be possible to build UCN-source providing a 100 times increased UCN density available for experiments compared to today's standards\,\cite{Kirch2014FPUA}.

\subsection{Systematic effects}
A substantial increase in counting statistics will immediately require to also control all systematic effects on a similar level. Two different types of systematic effects might be identified: i)~effects correlated to the change of the field configuration, i.e.\ a leakage current, mimicking the signal of an nEDM; and ii)~effects of stochastic nature, i.e.\ random magnetic field drifts, which reduce the attainable sensitivity. This becomes visible from the last line of equation\,(\ref{eq:DiffConfig}):

\begin{equation}
	\dn =  \frac{\hbar\left(\omega^{\parallel}-\omega^{\nparallel} \right)-2\mu_\text{n}\left(B^{\parallel}-B^{\nparallel}\right)}{2\left(E^{\parallel}-E^{\nparallel}\right)},
\label{eq:DiffConfig2}
\end{equation}

\noindent with $B^{\parallel} = B_0+\delta B^{\parallel} + B(E^{\parallel})$ not necessarily  equal to $B^{\nparallel} = B_0+\delta B^{\nparallel} + B(E^{\nparallel})$. An example of a direct correlated systematic effect is the magnetic field $B_\text{L}$ from a leakage current across the insulator. 
In a worst case scenario, where the current $I_{\rm L}$ flows on a circular path once around the cylindrical storage cell, the effect is $\dn^\text{false} \leq I_{\rm L}\cdot \unit[\pow{1}{-28}]{\ecm/nA}$. 
Additional magnetometers are used to safeguard against changes of the magnetic field between measurements. This requires a magnetometer concept with an accuracy better than:

\begin{equation}
	\sigma(B_\text{mag}) \leq 1/4 \frac{2\left|E\right|\sigma(\dn)}{\mu_\text{n}} 
	 =  \frac{\hbar}{4\mu_\text{n}\alpha T\!\sqrt{N}}
\label{eq:SensMagn}
\end{equation}

\noindent where the factor $1/4$ guarantees that the statistical sensitivity from UCN is not compromised by more than $\sim 10\%$. The proportionality constant for a $\unit[10]{kV/cm}$ field is $\mu_\text{n}/(2\left|E\right|) = \unit[7.7\!\times\!10^{-29}]{\ecm/fT}$. This means $\sigma(B_\text{mag})\leq \SI{30}{fT}$ for a statistical sensitivity of $\unit[\pow{1}{-26}]{\ecm}$ per measurement, a value typical aimed for by many new searches.
\subsubsection{Co-habiting magnetometers}
In measurements using only one precession chamber it is of paramount importance to be able to correct for any magnetic-field drift, while in a double chamber geometry it is important to correct for drifts of the magnetic-field gradient. The perfect solution for this problem are magnetically susceptible atoms which co-occupy the same volume as the neutrons. Three different atomic-isotope species, all spin-1/2, are typically considered for this task: \tHe, \magHg{} and \magXe. While \magHg{} is already employed\,\cite{Green1998} the other two are proposed for spectrometers in the future. 

Although these magnetometers allow for a measurement-to-measurement correction of the neutron precession frequency, they pose a certain risk of transferring systematic effects to the neutron measurement. In particular the geometric phase effect of spin-1/2 particles exposed to an inhomogenous magnetic field with vertical gradient $g_z$ and an electric field\,\cite{Pendlebury2004,Afach2015EPJA}, creates a correlated systematic effect of the order of the current statistical sensitivity. In the case of \magHg\ the transferred effect to the neutron is:

\begin{equation}
	d_{\rm Hg\rightarrow n}^{\rm false} = g_z \cdot \unit[\pow{4.4}{-27}]{\ecm \frac{cm}{pT}}.
\label{eq:falseEDMHg}
\end{equation}

\noindent While the direct effect from the neutron is still negligible:

\begin{equation}
	\dn^{\rm false} = g_z \cdot \unit[\pow{1.5}{-29}]{\ecm \frac{cm}{pT}}.
\label{eq:falseEDMn}
\end{equation}

The transferred effect can be mitigated for current experiments by measuring \dn\ as a function of 

\begin{equation}
	R = \frac{\omega_{\rm n}}{\omega_{\rm Hg}}=\frac{\gamma_{\rm n}}{\gamma_{\rm Hg}}\left(1+\frac{g_z\cdot h_{\rm CM}}{B}+\dots\right),
\label{eq:R}
\end{equation}

\noindent where $h_{\rm CM}$ is the center-of-mass offset between the UCN and the mercury ensemble. The true value is the crossing point of two curves found by reversing $B$. This is one example of the subtle differences in how UCN and thermal atoms sample the precession volume which lead to tiny magnetic-field dependent deviation of the frequencies of the neutron and the co-habiting isotope (for \magHg{} see Ref.\,\cite{Afach2014PLB}).

\subsubsection{Auxiliary magnetometers}
Local magnetometers placed outside of the precession chamber but still inside a magnetic shield are also often proposed or used to measure the magnetic field. They have the clear advantage not to see the electric field and hence should be free of an electric-field correlated effect. In Ref.\,\cite{Afach2014PLB} optical pump cesium magnetometers (Cs-OPM) were successfully used to decompose the magnetic field of the precession chamber into spherical polynomial harmonics to extract the vertical magnetic-field gradient $g_z$. Such techniques, especially when using vector-magnetometers\,\cite{Afach2015OExpress}, could proof useful in the future to correct for systematic effects, i.e.\ determine $g_z$ in equations\,(\ref{eq:falseEDMHg}, \ref{eq:falseEDMn}). 
An important implementation of less sensitive magnetometers is their use in systems compensating for the Earth's magnetic field and actively controlling coils to correct for magnetic field changes provoked by nearby magnets\,\cite{Afach2014JAP}.

% Head 2
\section{WORLD-WIDE EFFORTS SEARCHING FOR AN nEDM } 
\begin{figure}%
\centering
  \includegraphics[width=0.7\columnwidth]{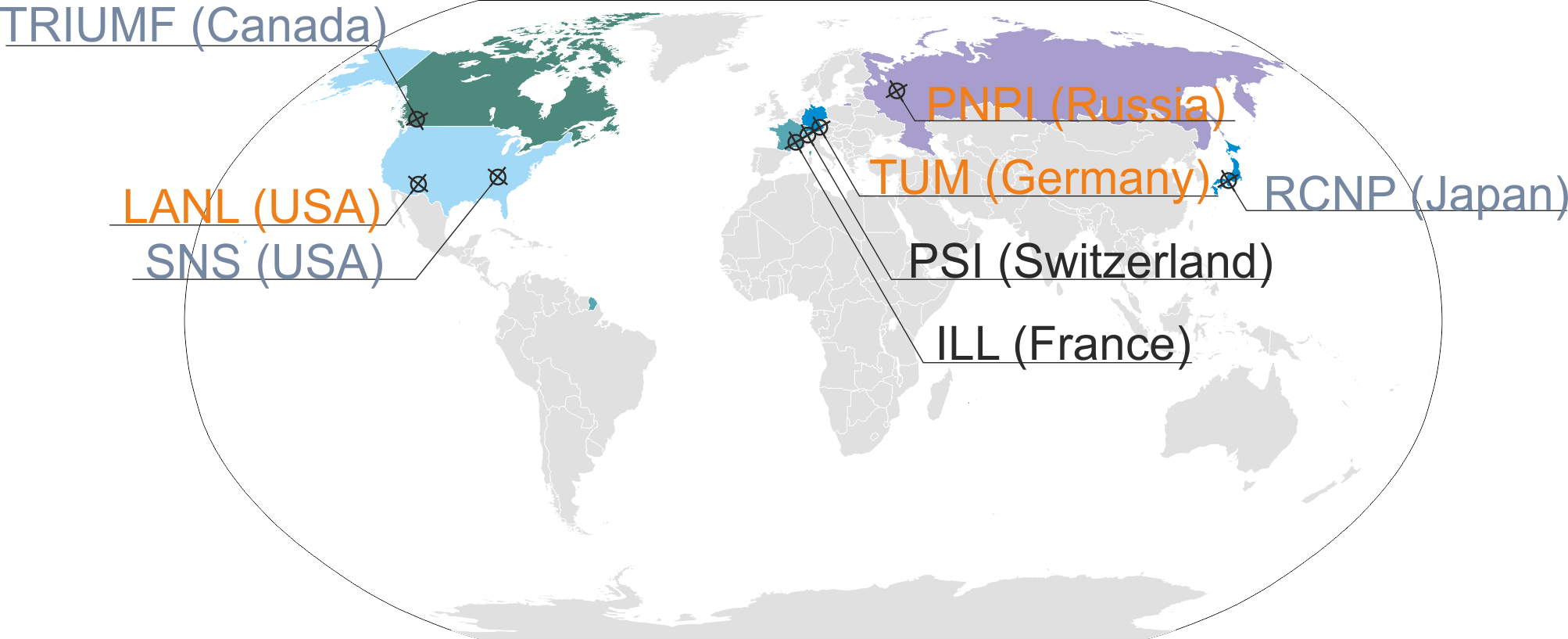}%
\caption{World map of nEDM searches.}%
\label{fig:WMapnEDM}%
\end{figure}

Several groups word wide, see Fig.\,\ref{fig:WMapnEDM}, compete in the effort to search for an electric dipole moment of the neutron. A detailed  summary of the individual projects can be found in Tab.\,\ref{tab:WWSearches} and in the references therein.

Currently only two groups/collaborations are in the formidable situation of being able to take data. Both groups, the nEDM collaboration at PSI, Switzerland and the PNPI UCN nEDM at ILL, France,  are aiming in the next year or two for a modest improvement of the current sensitivity into the low $\unit[10^{-26}]{\ecm}$ range. As these efforts are essentially limited by counting statistics the groups already work on or plan upgrades either by constructing a new spectrometer which is better adapted to the UCN source (PSI), or by moving to a better still to be constructed better source (ILL $\rightarrow$ PNPI).
All next generation designs aim for a sensitivity in the low $10^{-27}$ or even in the $\unit[10^{-28}]{\ecm}$ range. The concepts rely on new UCN source based on superthermal conversion, either using superfluid helium or solid deuterium, promising to deliver at least two orders of magnitudes more UCN\@. 

Two other projects stand out as they are not using UCN, but cold neutrons and their distinct techniques have not been discussed in theses proceedings. The crystal-diffraction nEDM at ILL, which is currently being upgraded, uses the very high electric field of non-center symmetric crystals\,\cite{Fedorov2011PhyB}. Whereas the project in discussion for the European spallations source at Lund, is proposing a double chamber beam experiment. The pulsed beam of the spallation source lends itself ideally to measure the $\mathbf{v}\times\mathbf{E}$-effect as a function of velocity and extrapolate to zero velocity for the real nEDM value\,\cite{Piegsa2013PRC}. 

\begin{table}%
\centering
\begin{tabular}{lcrclr}
\hline
	\bf Facility, location &\bf UCN (CN) source &\bf Spectrometer &\bf Magnetometry &\bf Status &\bf Sensitivity \\
	 &  & Reference &  & & $\mathcal{O}(\unit[10^{-27}]{\ecm}$)\\
	\hline
	LANL, USA & sD$_2$\,\cite{Saunders2013RSI} & dbl. chamber & \magHg & d/c & 1 \\
	ILL, France & cold source\cite{Ageron1978},  & dbl.\ chamber\,\cite{Serebrov2015PRC} & Cs-OPM & io & 10 \\
							&turbine\,\cite{Steyerl1986PLA}& & &  \\
  ILL, France & He-II to vacuum\,\cite{Zimmer2015PRC} & dbl. chamber\,\cite{Altarev2012a} & \magHg, Cs-OPM & id & 1\\
	ILL, France & CN beam\,\cite{Abele2006} & crystal-diffraction\,\cite{Fedorov2011PhyB} & - & u & 10 \\
	ESS, Sweden & pulsed CN & dbl.\ chamber\,\cite{Piegsa2013PRC}& - & id & 0.1 -10 \\
	FRM-II, Germany & sD$_2$\,\cite{Trinks2000NIMA} & dbl.\ chamber\,\cite{Altarev2012a}& \magHg, Cs-OPM & c  & 0.1 \\
	SNS, Oakridge & He-II in-situ\,\cite{Kolarkar2010AIP} & dbl.\ chamber\,\cite{Kolarkar2010AIP} & \tHe{}, squids & ccd & 0.1 \\
	PNPI, Russia& He-II to vacuum\,\cite{Serebrov2014TPL} & dbl.\ chamber\,\cite{Serebrov2015PRC} & Cs-OPM & c & 0.1-1 \\ 
	TRIUMF, Canada & He-II to vacuum\,\cite{Masuda2015JPS} & dbl.\ chamber\,\cite{Matsuta2013AIP} & \magHg, \magXe{}& d/c & 0.1-1 \\
	PSI, Switzerland & sD$_2$\,\cite{Lauss2014} & sgl./dbl.\ chamber\,\cite{Baker2011}& \magHg, \tHe, Cs-OPM &io/u & 1-10\\
	
\end{tabular}
\caption{Overview of all current world wide nEDM-projects. The status is indicated by: (id)~in discussion, (d)~design, (ccd)~critical components demonstration, (c)~construction, (o)~operation and (u)~uprading. }
\label{tab:WWSearches}
\end{table}

%
%\subsection{LANL - Los Alamos, United States of America}
%The UCN source of the Los Alamos Neutron Laboratory LANL is being currently upgraded. As part of the lab research activities a Ramsey type storage cell experiment at room temperature is planned which shall profit from the newly installed UCN-source by summer 2016. The initial goal is to demonstrate the ability to store sufficient UCN for a $\unit[\pow{1}{-26}]{\ecm}$ nEDM experiment.
%
%\subsection{FRM\,II - Munich, Germany}
%\subsection{SNS - Oakridge, United States of America}
%\subsection{PNPI - St.\ Petersbourg, Russia}
%\subsection{TRIUMF - Vancouver, Canada}
%\subsection{PSI - Villigen, Switzerland} 

\clearpage

% Head 3
\section{CONCLUSION}
World wide several groups are pursuing promising approaches to improve the sensitivity for the search of an electric dipole moment. In the next years most probably a new result will improve the current limit on the nEDM slightly, while an order of magnitude improvement seems likely in the next 5 years. Another order of magnitude will require a breakthrough in UCN counting statistics or a dramatic improvement of the electric field which seems to be possible in superfluid helium.

% Sections that will go in second font

% Acknowledgement
\section{ACKNOWLEDGMENTS}
I would like to thank all the colleagues who kindly provided input the for this overview presentation at the LASNPA-II in Medellin: V.~Fedorov, B.~Filippone, J.~Martin, A.~Serebrove and T.~Ito. Further I am grateful for all the discussion and feedback I had from the UCN group at PSI and the colleagues of the nEDM-collaboration.

Last but not least I would like to thank the organizers, especially Luis Fernando Cristancho, for their kind hospitality and the excellent conference giving me the opportunity to meet the Latin American nuclear physics community.
\def\aj{\ref@jnl{AJ}}                   % Astronomical Journal
\def\actaa{\ref@jnl{Acta Astron.}}      % Acta Astronomica
\def\araa{\ref@jnl{ARA\&A}}             % Annual Review of Astron and Astrophys
\def\apj{\ref@jnl{ApJ}}                 % Astrophysical Journal
\def\apjl{\ref@jnl{ApJ}}                % Astrophysical Journal, Letters
\def\apjs{\ref@jnl{ApJS}}               % Astrophysical Journal, Supplement
\def\ao{\ref@jnl{Appl.~Opt.}}           % Applied Optics
\def\apss{\ref@jnl{Ap\&SS}}             % Astrophysics and Space Science
\def\aap{\ref@jnl{A\&A}}                % Astronomy and Astrophysics
\def\aapr{\ref@jnl{A\&A~Rev.}}          % Astronomy and Astrophysics Reviews
\def\aaps{\ref@jnl{A\&AS}}              % Astronomy and Astrophysics, Supplement
\def\azh{\ref@jnl{AZh}}                 % Astronomicheskii Zhurnal
\def\baas{\ref@jnl{BAAS}}               % Bulletin of the AAS
\def\bac{\ref@jnl{Bull. astr. Inst. Czechosl.}}
                % Bulletin of the Astronomical Institutes of Czechoslovakia 
\def\caa{\ref@jnl{Chinese Astron. Astrophys.}}
                % Chinese Astronomy and Astrophysics
\def\cjaa{\ref@jnl{Chinese J. Astron. Astrophys.}}
                % Chinese Journal of Astronomy and Astrophysics
\def\icarus{\ref@jnl{Icarus}}           % Icarus
\def\jcap{\ref@jnl{J. Cosmology Astropart. Phys.}}
                % Journal of Cosmology and Astroparticle Physics
\def\jrasc{\ref@jnl{JRASC}}             % Journal of the RAS of Canada
\def\memras{\ref@jnl{MmRAS}}            % Memoirs of the RAS
\def\mnras{\ref@jnl{MNRAS}}             % Monthly Notices of the RAS
\def\na{\ref@jnl{New A}}                % New Astronomy
\def\nar{\ref@jnl{New A Rev.}}          % New Astronomy Review
\def\pra{\ref@jnl{Phys.~Rev.~A}}        % Physical Review A: General Physics
\def\prb{\ref@jnl{Phys.~Rev.~B}}        % Physical Review B: Solid State
\def\prc{\ref@jnl{Phys.~Rev.~C}}        % Physical Review C
\def\prd{\ref@jnl{Phys.~Rev.~D}}        % Physical Review D
\def\pre{\ref@jnl{Phys.~Rev.~E}}        % Physical Review E
\def\prl{\ref@jnl{Phys.~Rev.~Lett.}}    % Physical Review Letters
\def\pasa{\ref@jnl{PASA}}               % Publications of the Astron. Soc. of Australia
\def\pasp{\ref@jnl{PASP}}               % Publications of the ASP
\def\pasj{\ref@jnl{PASJ}}               % Publications of the ASJ
\def\rmxaa{\ref@jnl{Rev. Mexicana Astron. Astrofis.}}%
                % Revista Mexicana de Astronomia y Astrofisica
\def\qjras{\ref@jnl{QJRAS}}             % Quarterly Journal of the RAS
\def\skytel{\ref@jnl{S\&T}}             % Sky and Telescope
\def\solphys{\ref@jnl{Sol.~Phys.}}      % Solar Physics
\def\sovast{\ref@jnl{Soviet~Ast.}}      % Soviet Astronomy
\def\ssr{\ref@jnl{Space~Sci.~Rev.}}     % Space Science Reviews
\def\zap{\ref@jnl{ZAp}}                 % Zeitschrift fuer Astrophysik
\def\nat{\ref@jnl{Nature}}              % Nature
\def\iaucirc{\ref@jnl{IAU~Circ.}}       % IAU Cirulars
\def\aplett{\ref@jnl{Astrophys.~Lett.}} % Astrophysics Letters
\def\apspr{\ref@jnl{Astrophys.~Space~Phys.~Res.}}
                % Astrophysics Space Physics Research
\def\bain{\ref@jnl{Bull.~Astron.~Inst.~Netherlands}} 
                % Bulletin Astronomical Institute of the Netherlands
\def\fcp{\ref@jnl{Fund.~Cosmic~Phys.}}  % Fundamental Cosmic Physics
\def\gca{\ref@jnl{Geochim.~Cosmochim.~Acta}}   % Geochimica Cosmochimica Acta
\def\grl{\ref@jnl{Geophys.~Res.~Lett.}} % Geophysics Research Letters
\def\jcp{\ref@jnl{J.~Chem.~Phys.}}      % Journal of Chemical Physics
\def\jgr{\ref@jnl{J.~Geophys.~Res.}}    % Journal of Geophysics Research
\def\jqsrt{\ref@jnl{J.~Quant.~Spec.~Radiat.~Transf.}}
                % Journal of Quantitiative Spectroscopy and Radiative Transfer
\def\memsai{\ref@jnl{Mem.~Soc.~Astron.~Italiana}}
                % Mem. Societa Astronomica Italiana
\def\nphysa{\ref@jnl{Nucl.~Phys.~A}}   % Nuclear Physics A
\def\physrep{\ref@jnl{Phys.~Rep.}}   % Physics Reports
\def\physscr{\ref@jnl{Phys.~Scr}}   % Physica Scripta
\def\planss{\ref@jnl{Planet.~Space~Sci.}}   % Planetary Space Science
\def\procspie{\ref@jnl{Proc.~SPIE}}   % Proceedings of the SPIE

\let\astap=\aap
\let\apjlett=\apjl
\let\apjsupp=\apjs
\let\applopt=\ao

%\bibliography{nEDM-references,UCN-references,SM-references,BSM-references}%
%Control: author (3) initials jnrlst
%Control: year  (1) truncated

\end{document}